# Electronic manipulation of polar order in electron crystal

Takashi Kikkawa[1,‡,†], Ziyan Chen[1,†], Yuto Fujimoto[1], Takahiro Morimoto[1], Yuya Hirata[1], Hiroki Arisawa[1,2], Alexey A. Kaverzin[1], Satoshi Okamoto[3], Yoichi Okimoto[4], Naoshi Ikeda[5], and Eiji Saitoh[1,2,6,7]

*1. Department of Applied Physics, The University of Tokyo, Tokyo 113-8656, Japan.*

*2. RIKEN Center for Emergent Matter Science, Wako 351-0198, Japan.*

*3. Sumitomo Chemical Next-Generation Eco-Friendly Devices Collaborative Research Cluster, Kanagawa 226-8503, Japan.*

*4. Department of Chemistry, Institute of Science Tokyo, 2-12-1, Tokyo 152-8551, Japan.*

*5. Department of Physics, Okayama University, Okayama 700-8530, Japan.*

*6. Institute for AI and Beyond, The University of Tokyo, Tokyo 113-8656, Japan.*

*7. WPI Advanced Institute for Materials Research, Tohoku University, Sendai 980-8577, Japan.*

[‡] Present address: *Advanced Science Research Center, Japan Atomic Energy Agency, Tokai 319-1195, Japan.*

[†] These authors contributed equally: Takashi Kikkawa, Ziyan Chen.

## Abstract

**When interaction among atoms or ions is strong enough, they often arrange periodically, forming a crystal. The arrangement patterns of atoms or ions can encode information, a concept that has enabled devices such as ferroelectric memories. It has been found that not only atoms or ions but also electrons in condensed matter can crystallize when Coulomb interaction is strong enough. Typical examples are charge-ordered states in solids, where different valences, or different electron numbers, of an ion spontaneously form a spatial pattern on the lattice. In such electron crystals, information is expected to be encoded into the electron-ordering patterns. Here, we demonstrate electronic manipulation and readout of charge-ordering directions in a paramagnetic semiconductor $LuFe_2O_4$. By applying current pulses at room temperature, we observed that the non-reciprocal resistivity of $LuFe_2O_4$ is modulated along with a sign reversal, which disappears above the charge-ordering temperature. A numerical calculation incorporating inter-band Berry curvature affected by the charge ordering is consistent with the experimental results. By applying the observed phenomenon, we also demonstrate a non-reciprocal resistance memory operation in the charge-ordered $LuFe_2O_4$. This result opens the door to realizing charge-ordering electronics.**

## Main

$LuFe_2O_4$ is a typical charge-ordering system[1–6], in which Fe ions spontaneously arrange into a spatially patterned $Fe^{2+}$ and $Fe^{3+}$ below $T_{CO}$ ~ 320 K (refs.[7–13]), forming a charge superlattice on the crystal lattice. As shown in Fig. 1**a**, this $Fe^{2+}$–$Fe^{3+}$ pattern emerges on the doubly-stacked

Fe hexagonal layers along the *c* axis in $LuFe_2O_4$, resulting in alternating $Fe^{2+}$-rich and $Fe^{3+}$-rich layers[7,11–14] (see Figs. 1**a** and 1**b**). This $Fe^{2+}$–$Fe^{3+}$ arrangement exhibits valence-charge polarity, which is characterized by the polar order parameter **Π** normal to these double Fe layers[12,13] (see Fig. 1**a**). This order occurs without accompanying magnetic ordering.

In the absence of conduction electrons, this valence-charge polar ordering would cause an electric polarization, **P** (|| **Π**), and $LuFe_2O_4$ would become a ferroelectric[3,5,15–18]. However, the band gap of $LuFe_2O_4$ is very narrow ~ 0.1 eV (ref.[8]) and dense conduction electrons of ~ $10^{15}$ $cm^{-3}$ (ref.[10]) are present at least at room temperature in $LuFe_2O_4$; **P** is well screened and vanishes[17–22], and the system becomes a semiconductor. Therefore, neither **Π** nor **P** come up to the surface in the conventional electric properties of $LuFe_2O_4$. This situation impedes the control of the polar order **Π** in $LuFe_2O_4$, in contrast to conventional ferroelectrics[15,16] where **P** (and **Π**) can be switched by external electric fields **E** (Fig. 1**d**).

In spintronics[23], on the other hand, a new principle to control magnetization order parameter in conductors has been discovered, called spin-orbit torque (SOT)[24,25] (Fig. 1**e**). In SOT, electrons flowing through a device interact directly with the magnetization **M** and drive its reversal[26–28] (Fig. 1**e**), a concept which upends the conventional **M** control based on external magnetic fields **H** in magnetic memory devices[29] (Fig. 1**c**). Microscopically, the applied current or electric field **E** modulates electronic structure to generate electronic spin polarization, and drives the changes in **M** (refs.[26,28,30]). This effect can be interpreted as an appearance of "fictitious" magnetic field, $\mathbf{H}_{\mathrm{eff}}$ (refs.[26,28,30]) (Fig. 1**e**), which represents the modulation of **M** caused by the electronic-structure reconfiguration.

Such electronic control requires dense electronic states around the Fermi energy, in general, which allows a large number of electrons to participate in the reconfiguration. Metals or narrow-gap semiconductors are thus promising candidates. Furthermore, not only magnetization, this principle can be generalized to other order parameters in principle; the charge ordering in a narrow-gap semiconductor $LuFe_2O_4$ is expected to be manipulated by the similar electronic mechanism.

Here, we report current-induced manipulation of charge-ordering polarity and its electrical detection in $LuFe_2O_4$ at room temperature (Figs. 1**f** and 2**a**). Theoretical calculation reflecting charge-order modulation well reproduces the experimental results. The effect can be regarded as a charge-dipole analogue of the current-induced spin polarization (SOT) and resulting magnetization manipulation discussed in spintronics[26,28,30] (Figs. 1**e** and 1**f**). Our finding provides an effective method for controlling and detecting various orders in conductors.

**Experimental procedure**

To detect the electronic control of the charge-ordering polarity, we have used second-order nonlinear resistance measurement. In the charge-ordered phase of $LuFe_2O_4$, a charge superstructure of $Fe^{2+}$-rich and $Fe^{3+}$-rich layers forms along the *c* axis (Fig. 1**a**), along which the two opposite directions are distinguishable due to the difference in the layer distances in the unit cell[7,12,13]. This means the polar symmetry breaking down along the *c* axis, although the corresponding charge polarization is almost screened by the conduction electrons around room temperature[22]. If conduction electrons can feel this atomic polar ordering, the resistance $R$ may differ between the opposing currents of $+I$ and $-I$, a phenomenon known as the non-reciprocity: $R(+I) \neq R(-I)$ (refs.[31,32]). In this case, the voltage $V$ component proportional to $I^2$ appears: $V = R_{1\omega}I + R_{2\omega}I^2 \cdots$, where $R_{1\omega}$ represents the conventional linear resistance, whereas the non-reciprocity is transcribed into the nonlinear resistance $R_{2\omega}$. The nonlinear term $V_{2\omega} = R_{2\omega}I^2$ can be precisely measured by a lock-in method[31–33].

To experimentally control and detect the charge ordering in $LuFe_2O_4$, we employed a lock-in technique that combines pulsed-d.c. and a.c. currents (see Figs. 2**a** and 2**b**). When an a.c. current, $I_{ac} = \sqrt{2}I_{rms}\sin\omega t$, is applied to the sample, the nonlinear voltage ($\propto I_{ac}^2$) is generated at the doubled frequency $2\omega$ since $I_{ac}^2 \propto \sin^2 \omega t \propto (1 - \cos 2\omega t)$. This voltage can be detected as a second harmonic ($2\omega$) voltage $V_{2\omega}$ using a lock-in amplifier[33]. We also apply a d.c. current to manipulate the charge-ordering polarity $\mathbf{\Pi}$ (Fig. 2**a**). In the measurement, contribution of thermoelectric voltage $V_{2\omega}$ due to heating[34–36] is excluded by utilizing pulsed d.c.-current; we first apply a square-pulsed d.c. current, $I_{pulse}$, for the $\mathbf{\Pi}$ control ($|I_{pulse}| \leq 2$ mA), followed by waiting for a brief quarter-minute, and measure the $2\omega$ lock-in voltage using an a.c. sensing current, $I_{ac} = \sqrt{2}I_{rms}\sin\omega t$ (with $|\sqrt{2}I_{rms}| \leq 100$ μA). All measurements were performed under the standard four-terminal geometry to avoid possible contribution from electrodes (see Figs. 2**a** and 2**b**). This allows precise measurement of non-reciprocal response influenced by the charge-order patterns controlled by $I_{pulse}$. We used two rectangular-shaped $LuFe_2O_4$ samples with different dimensions, labeled as 1 and 2, having the resistance of $R$ = 9.0 and 1.4 kΩ, respectively (see Methods).

**Observation of electronic-induced polar order switching**

In Figs. 2**c** and 2**d**, we show the voltage $V_{2\omega}$ data for the $LuFe_2O_4$ sample 1 at room temperature,

measured while sweeping the pulsed current $I_{\mathrm{pulse}}$. When the current is applied along the charge-ordered planes (i.e., $I_{\mathrm{ac}}$, $I_{\mathrm{pulse}} \perp$ $c$-axis), no signal is observed (see Fig. 2**c**). By stark contrast, when the current is applied across the charge-ordered planes (i.e., $I_{\mathrm{ac}}$, $I_{\mathrm{pulse}} \parallel$ $c$-axis), an unconventional $V_{2\omega}$ signal clearly shows up in the $LuFe_2O_4$ sample (see Fig. 2**d**). When starting with a current pulse of $I_{\mathrm{pulse}}$ = +1.8 mA applied in the $+z$ direction in Fig. 2**b**, a positive $V_{2\omega}$ signal appears. Then, as the amplitude of $I_{\mathrm{pulse}}$ is gradually reduced, the positive $V_{2\omega}$ signal persists even when the $I_{\mathrm{pulse}}$ reaches zero (see red circles in Fig. 2**d**). Subsequently, when the current-pulse polarity is reversed to the $-z$ direction (i.e., $I_{\mathrm{pulse}} < 0$ in Fig. 2**d**), $V_{2\omega}$ gradually decreases with increasing $|I_{\mathrm{pulse}}|$, reaching zero at $I_{\mathrm{pulse}} \sim -0.7$ mA. Further increasing the pulse amplitude in the $-z$ direction reverses the sign of $V_{2\omega}$. As $I_{\mathrm{pulse}}$ decreases toward $I_{\mathrm{pulse}}$ = –1.8 mA, $V_{2\omega}$ becomes saturated again for $I_{\mathrm{pulse}} \lesssim -1.2$ mA. Conversely, when sweeping $I_{\mathrm{pulse}}$ from –1.8 mA to +1.8 mA (orange circles in Fig. 2**d**), the initially negative $V_{2\omega}$ signal remains saturated, but gradually increases as the current-pulse polarity is reversed ($I_{\mathrm{pulse}} > 0$), reaching zero around +0.7 mA and turning positive as $I_{\mathrm{pulse}}$ is further increased. For $I_{\mathrm{pulse}} \gtrsim$ 1.2 mA, the $V_{2\omega}$ signal eventually almost saturates. The hysteresis behavior of $V_{2\omega}$ is reminiscent of a *P–E* curve in conventional ferroelectrics[15], although no *P–E* signal can be measured for the present $LuFe_2O_4$ at room temperature[22]. As shown in Figs. 3**c** and 3**d**, we found that the signal magnitude increases in proportion to the squared $I_{\mathrm{ac}}$ amplitude ($V_{2\omega} \propto I_{\mathrm{rms}}^2$), confirming that the observed $2\omega$ signal is due to the second-order nonlinear voltage reflecting the non-reciprocal electron transport[31–33]. We also confirmed similar $V_{2\omega}$ versus $I_{\mathrm{pulse}}$ characteristics for another $LuFe_2O_4$ sample (sample 2, Fig. 3**b**).

Importantly, the signal disappears above the charge-ordering transition temperature ($T_{CO}$ = 320 K) of $LuFe_2O_4$, as shown in Fig. 3**a**. The result shows that the charge ordering is indispensable for the observed signal to appear. The observed current-direction dependence and the disappearance of the signal above $T_{CO}$ cannot be explained by other possible nonlinear effects, such as mass transport and extrinsic interfacial effects, as well as thermoelectric effects.

From the data shown in Figs. 2**d** and 3**a**, the threshold pulsed current $I_{th}$ at which $V_{2\omega}$ becomes zero is estimated to be $I_{th}$ = 0.72 mA and 1.15 mA for the sample 1 and 2, corresponding to the current density of $j_{th} = 3.8\times10^{-2}$ Acm$^{-2}$ and $2.3\times10^{-2}$ Acm$^{-2}$, respectively. These values represent the minimum current required to switch the charge-order polarity in $LuFe_2O_4$, which are several orders of magnitude smaller than that required for typical current-induced magnetization switching in spintronic devices[24,25,37–40]. Besides, converting the $I_{th}$ values to coercive electric fields $E_c$ (= $RI_{th}/d$) yields $E_c$ ~ 73 Vcm$^{-1}$ and 25 Vcm$^{-1}$ for the sample 1 and 2, respectively ($d$ is the distance between the electrodes). These values are remarkably small compared to those typically observed in conventional ferroelectrics[15], showing that the present switching is achieved with much smaller power consumption. From the data, we estimate the nonlinear transport coefficient, defined as $\gamma \equiv E_{2\omega}/(\rho j_{\mathrm{rms}}^2)$, to be $3.1\times10^{-7}$ A$^{-1}$m$^2$ and $5.3\times10^{-7}$ A$^{-1}$m$^2$ for the sample 1 and 2, respectively, around $I_{pulse}$ = 0 mA at room temperature. Here, $E_{2\omega}$ is the second harmonic voltage normalized by the electrode length $d$, $\rho$ is the resistivity of $LuFe_2O_4$ ($1.9\times10^2$ and $1.0\times10^2$ Ωcm for the sample 1 and 2, respectively), $j_{rms}$ is the root-mean-squared current density. These values are comparable between the two samples.

**Demonstration of non-volatile memory function**

We demonstrate memory function operation by writing information into charge order using the observed hysteresis feature. To this end, we applied pulsed currents in $LuFe_2O_4$ in a sequential positive-to-negative pattern and measured the $2\omega$ voltage $V_{2\omega}$ induced by an a.c. current at room temperature. As shown in Fig. 4, $V_{2\omega}$ exhibits a similar alternating pattern, and even after turning off the pulsed current, the $V_{2\omega}$ maintains similar values until the next pulse is activated. This result demonstrates a non-volatile memory function, enabling charge-order memory.

**Theoretical approach for the electronic-induced polar-order writing and reading**

To understand the observed charge-order writing and reading, we construct a minimal theoretical model for electronic reconfiguration and charge-order modulation induced by the applied electric currents or fields (see Methods for the details of the model and calculation). Specifically, we model the $Fe^{2+}$ and $Fe^{3+}$ layers in $LuFe_2O_4$ by a simple one-dimensional model called Rice–Mele model[41] (Fig. 5**a**) with two non-equivalent sites, $A$ and $B$, breaking the inversion symmetry (see Extended Data Fig. 2 and Fig. 5**b**). The charge ordering, characterized by the polar order parameter $\Pi$ (see Methods), is stabilized by inter-site Coulomb interaction, while the on-site Coulomb force leads to a staggered mean-field potential for the $A$ and $B$ sites depending on the direction of the charge ordering (see Extended Data Fig. 2).

Upon the application of an external electric field $E$, the $E$ induces inter-band hybridization[42], which can be strong when the band gap is small (see Fig. 5**a**). Assume that the chemical potential lies in the valence band $\varepsilon_{-}(k)$. The valence wave function undergoes inter-band hybridization with the conduction wave functions in the first-order perturbation of $E$. In the real space, this inter-band hybridization corresponds to the mixing of the electron orbitals between the $A$ and $B$ sites, resulting in a modification of the electron distribution between these sites. The electron reconfiguration is illustrated in Figs. 5**b** (for external $E = 0$) and 5**c** (for $E \neq 0$), where the red and blue curves schematically represent the electron distribution at the $A$ and $B$ sites, $\langle n_A \rangle$ and $\langle n_B \rangle$ (the size of the blue circle at each site also represents the amount of charge). The charge-distribution modulation, described as the change in the polar-order parameter $\delta\Pi$ from its equilibrium value $\Pi_0$, arises in proportion to $E$ as $\delta\Pi = \Pi - \Pi_0 = \xi E$. The coefficient $\xi$ has a geometrical meaning and is directly related to the Berry curvature $F_{k\Pi}$ (refs.[43,44]) in the parameter space spanned by the momentum $k$ and the polar order $\Pi$ (for details, see Methods). Since the Berry curvature $F_{k\Pi}$ quantifies the polarization current under the change of the parameter $\Pi$ in the theory of electric polarization[43,44], this expression indicates that the charge-order modulation can be interpreted as a type of polarization currents induced by the applied electric field. $\Pi_0 + \xi E$ in the non-equilibrium state modifies the free energy, $F_{\text{Coulomb}} = -V\Pi^2 + \mathcal{O}(\Pi^4) + \cdots$, by

$$\Delta F_{\text{Coulomb}} \sim -2V\xi E \cdot \Pi_0 \equiv -E_{\text{eff}} \cdot \Pi_0, \qquad (1)$$

where $V$ represents the inter-site Coulomb interaction. This can be regarded as emergence of an effective field $E_{\text{eff}} = 2V\xi E (\propto \delta\Pi)$, which couples with the polar order $\Pi_0$ upon the application of $E$. Importantly, this coupling far exceeds the conventional $P \cdot E$ coupling by a factor of $\sim 10^{3000}$, because the macroscopic electric polarization $P$ is screened by mobile electrons and almost vanishes in the present $LuFe_2O_4$ at room temperature. In contrast, the modulation of the Coulomb energy in the unit cell described in Eq. (1) occurs on the atomic scale and is not affected by such screening (see Methods). Under the application of $E$, the two minima of $F_{\mathrm{Coulomb}}$ (Fig. 5**d**) are asymmetrically shifted, favoring a particular direction of the charge-order polarity as depicted in Fig. 5**d**, allowing the manipulation of $\Pi$.

The modulation of the charge order under the application of the electric field, $\Pi_0 \rightarrow \Pi_0 + \xi E$, also impacts the electron band dispersion $\varepsilon_{-}(k)$ with respect to the $E$ direction, inducing $E$-dependent modulation of the electron group velocity (see Fig. 5**e** and Methods). Since the electric conductivity is proportional to the Fermi velocity, such modulation of the electron group velocity leads to a nonlinear current, $j^{(2)} \propto E^2$, even under the time reversal symmetric condition[45]. The mechanism can be responsible for the observed non-reciprocal transport and generation of second-order nonlinear voltage upon the application of a current.

The present effect can be applied to making memories utilizing charge-order degrees of freedom. It is thus surely worthwhile to explore this effect in various conductors in which inversion symmetry is broken by the presence of charge ordering.

## Methods

### Sample preparation.

$LuFe_2O_4$ single crystals were grown using a floating-zone (FZ) melting method. $Fe_2O_3$ (5N, Rare Metallic Co., Ltd.) and $Lu_2O_3$ (4N, Nippon Yttrium Co., Ltd.) powders were mixed and pressed into a rod shape for the FZ process with hydrostatic pressure. The crystals were grown at a rate of 5 $mmh^{-1}$ under a $CO/CO_2$ gas mixture. Further details on the sample-growth procedure are described in ref.[13].

Two $LuFe_2O_4$ samples, labeled as 1 and 2, were used for the transport measurement and obtained by dicing the $LuFe_2O_4$ single crystals. Sample 1 (2) has the length of 2.12 mm (0.95 mm) along the *c* axis and the cross section of 1.90 $mm^2$ (4.94 $mm^2$). The distance between the two-voltage electrodes along the *c* axis for sample 1 (2) is 0.89 mm (0.64 mm).

### Sample characterization by X-ray diffraction.

Extended Data Fig. 1 displays an X-ray diffraction pattern of the reciprocal lattice plane $[(H\ H\ 0)_h - (0\ 0\ L)_h]$ of the $LuFe_2O_4$ single-crystalline sample obtained by the X-ray oscillation photography. Clear diffraction spots are observed at superlattice positions such as $(1/3\ 1/3\ L)_h$ and $(2/3\ 2/3\ L)_h$ with $L = 0.5n$ ($n$: integer). The result confirms the presence of three-dimensional charge ordering and high crystallinity of our sample, as discussed in refs.[12,46,47].

### Nonlinear transport measurement.

We measured non-reciprocal electron transport affected by the charge ordering using a lock-in detection technique in a four-terminal geometry employing a micro-probing system under high vacuum conditions. Here, a standard four-probe resistance measurement setup with a constant a.c.-current amplitude was adopted. This setup eliminates spurious contribution from the electrical contacts, a key distinction from conventional two-terminal dielectric measurement. As an input, we apply two types of currents: A square-pulsed current $I_{pulse}$ to control the charge order and a.c. current $I_{ac}$ $(=\sqrt{2}I_{rms}\sin\omega t)$ to electrically read out the charge-order polarity, both parallel to the *c* axis of $LuFe_2O_4$ (except for the control experiment shown in Fig. 2**c**). Here, $\omega$, $t$, and $I_{rms}$ denote the angular frequency (corresponding to the frequency $f = \omega/2\pi$), time, and root-mean-square (rms) amplitude of the a.c. current, respectively. The nonlinear voltage ($\propto I_{ac}^2$) oscillates at the angular frequency of $2\omega$, and it is selectively measured as a second harmonic ($2\omega$) voltage using a lock-in method[31–33]. Both $I_{pulse}$ and $I_{ac}$ are generated from a same current source (6221, Keithley), and the $2\omega$ voltage $V_{2\omega}$ is measured using a lock-in amplifier (LI5650, NF Corporation). To minimize possible heating due to $I_{pulse}$, we introduce a typical interval of 15 seconds before acquiring the $V_{2\omega}$ data. The pulsed current has an amplitude of $|I_{pulse}| \leq 2$ mA with a width of 40–80 ms. The a.c. current has an $I_{ac}$ peak-amplitude ($=\sqrt{2}I_{rms}$) ranging from 2–100 μA and the frequency of 13 Hz. The data are anti-symmetrized with respect to $I_{pulse}$.

**Theoretical modeling based on minimal two-band model.**

In this section, we present a theoretical discussion on the origin of the $\Pi$-order manipulation and its electric readout via the nonlinear voltage. In the model elaborated below, the $\Pi$ manipulation is attributed to the electronic-induced charge redistribution (or charge polarization) stemming from inter-band hybridization, which can be regarded as a charge analogue of the current-induced spin polarization found in spintronics[26,28,30,48]. The $\Pi$ reading is attributed to the asymmetry of the conductivity with respect to the applied current direction resulting from the field-induced modulation of the electron band curvature (and hence, group velocity).

To model conduction electrons affected by the polar ordering of $Fe^{2+}$ and $Fe^{3+}$ ions in $LuFe_2O_4$ on a mean-field level, we consider a one-dimensional system consisting of two nonequivalent sites, $A$ and $B$, in a unit cell, separated by the distance $a$. In this model, itinerant electrons (hereafter referred to as $p$ electrons) experience potential, $U\Pi$, representing on-site Coulomb repulsion from the localized $d$ electrons of $Fe^{2+}$ or $Fe^{3+}$ located at the $A$ and $B$ sites (see Extended Data Fig. 2 and note that the Fe valence states are switchable between the $A$ and $B$ sites, corresponding to the switching of charge-ordering pattern, or the polar-order parameter $\Pi$). (See also the subsequent sections in Methods for further details on the order parameter $\Pi$ and its interplay with itinerant $p$ electrons.) Here, $U\Pi$ is the product of the polar order parameter $\Pi$ $(-1 \leq \Pi \leq -1)$ and the on-site mean-field Coulomb potential $U$. To account for the broken spatial inversion symmetry, we introduce alternating hopping parameters, denoted as $t + \delta t$ and $t - \delta t$ (see Extended Data Fig. 2**b**). Then, the Hamiltonian for itinerant electrons is given by

$$H = \sum_{i}^{N} \left[ \left( (t - \delta t) c_{iB}^{\dagger} c_{iA} + (t + \delta t) c_{iA}^{\dagger} c_{i-1B} + h.c. \right) - U\Pi c_{iA}^{\dagger} c_{iA} + U\Pi c_{iB}^{\dagger} c_{iB} \right], \qquad (2)$$

which is called the Rice–Mele Hamiltonian[41]. Here, $c_{iA(B)}^{\dagger}$ and $c_{iA(B)}$ are the creation and annihilation operators of electrons at the cell $i$, respectively, and $N$ denotes the number of unit cells. By the Fourier transform $c_{iA(B)} = \frac{1}{\sqrt{N}} \sum_k e^{ikr_i} c_{kA(B)}$ with $r_i$ being the real-space position of the cell $i$, we express the Hamiltonian as

$$H = \sum_k \left(c_{kA}^\dagger \, c_{kB}^\dagger\right) H_k \begin{pmatrix} c_{kA} \\ c_{kB} \end{pmatrix}, \tag{3}$$

where

$$H_k = \begin{pmatrix} -U\Pi & (t+\delta t)e^{-ika} + (t-\delta t)e^{ika} \\ (t+\delta t)e^{ika} + (t-\delta t)e^{-ika} & U\Pi \end{pmatrix}. \tag{4}$$

By diagonalization, we obtain the following two energy bands:

$$\mathcal{E}_\pm(k) = \pm|\mathbf{d}(k)| = \pm\sqrt{(U\Pi)^2 + 2t^2 + 2\delta t^2 + 2(t^2 - \delta t^2)\cos 2ka}, \tag{5}$$

as schematically depicted in Fig. 5**a**. Here, the higher and lower energy bands are denoted as $\mathcal{E}_+(k)$ and $\mathcal{E}_-(k)$, respectively, with the associated wavefunctions of $|+\rangle$ and $|-\rangle$ given by

$$|+\rangle = \begin{pmatrix} \cos\frac{\theta}{2} \\ e^{i\phi}\sin\frac{\theta}{2} \end{pmatrix}, \tag{6}$$

$$|-\rangle = \begin{pmatrix} e^{-i\phi}\sin\frac{\theta}{2} \\ -\cos\frac{\theta}{2} \end{pmatrix}, \tag{7}$$

with $\mathbf{d}(k) = (d_x, d_y, d_z) = (2t\cos ka, 2\delta t\sin ka, -U\Pi)$, $\cos\theta = d_z/|\mathbf{d}|$, and $\tan\phi = d_y/d_x$. In the following, without loss of generality, we consider the case where the chemical potential lies within the $\mathcal{E}_-$ band.

We apply the first-order perturbation theory upon the application of an electric field (generated by the voltage drop from the current due to the resistance) by introducing the perturbation $\hat{V} = -eE\hat{r}$, ($e(= -|e|)$: the electron charge, $\hat{r}$: the position operator of electrons), which leads to the correction of the wavefunction $|-\rangle$ into

$$\widetilde{|-\rangle} = |-\rangle - eE\frac{\langle +|\hat{r}|-\rangle}{\mathcal{E}_- - \mathcal{E}_+}|+\rangle. \tag{8}$$

The second term can be significantly large in narrow-gap systems. From the Heisenberg equation $\hat{v} = \left(\frac{1}{i\hbar}\right)[\hat{r}, H]$, the matrix element of $\hat{r}$ can be written with $\langle +|\hat{v}|-\rangle = (1/i\hbar)(\mathcal{E}_- - \mathcal{E}_+)\langle +|\hat{r}|-\rangle$, leading to

$$|\widetilde{-}\rangle = |-\rangle - i\hbar eE\frac{\langle +|\hat{v}|-\rangle}{(\mathcal{E}_- - \mathcal{E}_+)^2}|+\rangle. \tag{9}$$

Here the velocity operator is expressed as $\hat{v} = \frac{1}{\hbar}\partial_k H$.

The charge ordering $\Pi$ of localized $d$ electrons gives rise to the $p$-electron polarization $\Pi_{p,0}$ which is quantified by the difference in the average $p$-electron occupation between the $A$ and $B$ sites $\langle n_A^0\rangle - \langle n_B^0\rangle$,

$$\Pi_{p,0} = \langle n_A^0\rangle - \langle n_B^0\rangle = \frac{a}{\pi}\int \langle -|\sigma_z|-\rangle f_0 dk = \frac{a}{\pi}\int \frac{U\Pi_0}{|\mathbf{d}(k)|} f_0 dk, \tag{10}$$

where $\sigma_i$ is the Pauli matrix acting on the sublattice degrees of freedom and $f_0$ is the Fermi–Dirac distribution function (see Fig. 5**b**, where a schematic equilibrium charge occupation is illustrated, and see also its caption). Here, the quantities labeled with the sub- or super-script 0 indicate their values at equilibrium. Similarly, the non-equilibrium $p$-electron polarization $\Pi_p$ under $E$ can be calculated as the difference in the average occupation, $\langle n_A\rangle - \langle n_B\rangle$, influenced by the perturbation,

$$\Pi_p = \langle n_A\rangle - \langle n_B\rangle = \Pi_{p,0} + \delta\Pi_p, \tag{11}$$

$$\delta\Pi_p = \delta\langle n_A\rangle - \delta\langle n_B\rangle = \frac{a}{\pi}\int \left(\langle\widetilde{-}|\sigma_z|\widetilde{-}\rangle - \langle -|\sigma_z|-\rangle\right) f_0 dk. \tag{12}$$

Here, $\delta\langle n_A\rangle$ ($\delta\langle n_B\rangle$) denotes the modification of the electron occupation at the $A$ ($B$) site by $E$, as illustrated in Fig. 5**c**, and satisfies $\delta\langle n_A\rangle = -\delta\langle n_B\rangle$. Substituting Eq. (9) into Eq. (12) yields

$$\delta\Pi_p = -i\hbar eE\frac{a}{\pi}\int\frac{\langle -|\sigma_z|+\rangle\langle +|\hat{v}|-\rangle}{(\mathcal{E}_- - \mathcal{E}_+)^2}f_0 dk + h.c. = \xi_p E. \tag{13}$$

$\delta\Pi_p$ appears in proportion to $E$ [the expression of $\xi_p$ in Eq. (13) is given by Eq. (14)].

$\delta\Pi_p$ has a geometrical meaning related to the Berry curvature of the Bloch electrons. To see this, we rewrite the expression for $\xi_p$ as

$$\xi_p = -ie\frac{a}{\pi}\int\frac{\langle -|(-\partial_\Pi H/U)|+\rangle\langle +|\partial_k H|-\rangle}{(\mathcal{E}_- - \mathcal{E}_+)^2}f_0 dk + h.c.$$

$$= -\frac{e}{U}\frac{a}{\pi}\int 2\mathrm{Im}\left[\frac{\langle -|\partial_\Pi H|+\rangle\langle +|\partial_k H|-\rangle}{(\mathcal{E}_- - \mathcal{E}_+)^2}\right]f_0 dk. \tag{14}$$

Then we consider the Berry curvature $F_{k\Pi}$ defined in the parameter space $(k, \Pi)$, which is given by $F_{k\Pi} = \partial_k a_\Pi - \partial_\Pi a_k$ with the Berry connections $a_\lambda = i\langle -|\partial_\lambda|-\rangle$ ($\lambda = k, \Pi$) (refs.[43,44]). Since the Berry curvature $F_{k\Pi}$ can be expressed as

$$F_{k\Pi} = -2\mathrm{Im}\left[\frac{\langle -|\partial_k H|+\rangle\langle +|\partial_\Pi H|-\rangle}{(\mathcal{E}_- - \mathcal{E}_+)^2}\right], \tag{15}$$

in the case of two band models[49], the coefficient $\xi_p$ for the polar-order modulation is directly proportional to the Berry curvature as

$$\xi_p = \frac{e}{U}\frac{a}{\pi}\int F_{k\Pi} f_0 dk. \tag{16}$$

The expression means that the charge-order modulation can be interpreted as a kind of a polarization current quantified by the Berry curvature $F_{k\Pi}$.

The $d$-electron polar order $\Pi$ is also modified by the modulation of the $p$-electron polarization $\Pi_p$ through the on-site Coulomb interaction between $d$- and $p$-electrons. As detailed in the section "Microscopic perspective on the electronic manipulation of charge ordering" in Methods, the modulation of the $p$-electron polarization $\delta\Pi_p$ results in that of the $d$-electron polar order $\delta\Pi$ directly proportional to $\delta\Pi_p$, which we write as $\delta\Pi = \xi E$.

We here examine the impact of the electronic-induced $\delta\Pi$ on the free energy within our model. The free energy in the charge-ordered phase can be expanded as

$$F_{\mathrm{Coulomb}} = -V\Pi^2 + \beta\Pi^4 + \cdots, \tag{17}$$

where $V > 0$ and $\beta$ is a parameter associated with Coulomb interaction ($|\beta| < V$) [for the derivation of Eq. (17), see the section "Microscopic perspective on the electronic manipulation of charge ordering" in Methods]. In the absence of an external field, $F_{\mathrm{Coulomb}}$ displays two minima, corresponding to the spontaneous charge order of opposite orientations, which are degenerate in energy (see Fig. 5**d**). Upon the application of $E$, due to the non-equilibrium modulation of charge distribution, the order parameter $\Pi_0$ changes to $\Pi_0 + \xi E$ [Eq. (13)]. Substituting this into Eq. (17) yields the following term as the lowest-order contribution:

$$\Delta F_{\mathrm{Coulomb}} = -2V\xi E \cdot \Pi_0 = -E_{\mathrm{eff}} \cdot \Pi_0. \tag{18}$$

This term is significant due to the atomic Coulomb potential $V$, which are not well screened by the conduction electrons, making it orders of magnitude greater than the macroscopic dipolar coupling $P \cdot E$ if $P$ is screened by conduction electrons (see the next section in Methods). As evident in Eq. (18), $\Delta F_{\mathrm{Coulomb}}$ can be interpreted as the emergence of the effective field $E_{\mathrm{eff}} = 2V\xi E$ coupled with $\Pi_0$. Due to the energy shift $\Delta F_{\mathrm{Coulomb}} = -E_{\mathrm{eff}} \cdot \Pi_0 \propto E$, the two minima of the free energy are modulated asymmetrically depending on the $E$ direction (Fig. 5**d**). Under the sufficiently large $E$, stabilization of which sites of $A$ or $B$ is preferentially occupied, i.e., the orientation of the charge order, can be switched, as schematically illustrated in Fig. 5**d**. This enables the writing of the $\Pi$-order information.

Under an electric field, as discussed above, the order parameter $\Pi$ is modulated as $\Pi_0 + \xi E$.

In response, the energy difference between the $A$ and $B$ sites varies with the electric field as $U(\Pi_0 + \xi E)$, reflecting Hartree correction[45]. This results in a modification of the electron band curvature depending on the $E$ direction; by letting $\Pi = \Pi_0 + \xi E$ [Eq. (13)] in the electron band $\mathcal{E}_-(k)$ [Eq. (5)], we obtain

$$\mathcal{E}_-(k, E) = -\sqrt{U^2(\Pi_0 + \xi E)^2 + 2t^2 + 2\delta t^2 + 2(t^2 - \delta t^2)\cos 2ka}$$

$$\sim \mathcal{E}_-(k, E = 0) + \frac{U^2 \Pi_0 \xi}{\mathcal{E}_-(k, E = 0)} E, \quad (19)$$

where $\mathcal{E}_-(k, E = 0)$ is given by Eq. (5). The second term in Eq. (19) represents the modulation of the band curvature depending on the $E$ direction (see Fig. 5**e**). This causes a change of the electron group velocity $(1/\hbar)\partial\mathcal{E}_-(k, E)/\partial k|_{k=k_\mathrm{F}}$ asymmetric with respect to the $E$ direction ($k_\mathrm{F}$: the Fermi wavenumber ($k_\mathrm{F} > 0$)), leading to non-reciprocal electron transport[45]. Based on the Boltzmann transport theory under the relaxation-time ($\tau$) approximation, we obtain the following expression for a steady-state electric current up to the second order,

$$j = \int \left(\frac{e}{\hbar}\frac{\partial \mathcal{E}_-(k, E)}{\partial k}\right)\left(f_0 - \frac{e\tau}{\hbar}\frac{\partial f_0}{\partial k}E\right) dk = \sigma^{(1)}E + \sigma^{(2)}E^2, \quad (20)$$

with

$$\sigma^{(1)} = \frac{2e^2\tau}{\hbar^2}\frac{\partial \mathcal{E}_-(k)}{\partial k}\bigg|_{k=k_\mathrm{F}}, \quad (21)$$

$$\sigma^{(2)} = \frac{2e^2\tau}{\hbar^2} U^2 \Pi_0 \xi \frac{\partial}{\partial k}\frac{1}{\mathcal{E}_-(k)}\bigg|_{k=k_\mathrm{F}}. \quad (22)$$

The nonlinear conductivity $\sigma^{(2)}$ originates from the $E$-dependent term in Eq. (19). The $\sigma^{(2)}E^2$ term can be responsible for the non-reciprocal electron transport even under the time-reversal symmetric condition.

Finally, we give an order estimation of the non-reciprocal current for the present mechanism.

We consider the non-reciprocity ratio of the asymmetric part and the symmetric part of the voltage $\left|\frac{V(I)-V(-I)}{V(I)+V(-I)}\right| \simeq \left|\frac{V_{2\omega}}{V_{1\omega}}\right| \simeq \left|\frac{R_{2\omega}I}{R_{1\omega}}\right| \simeq \left|\frac{\sigma^{(2)}E}{\sigma^{(1)}}\right|$ ($V_{n\omega}$: $n$-th harmonic voltage). Using Eqs. (16), (21) and (22), this can be roughly evaluated as

$$\frac{\sigma^{(2)}E}{\sigma^{(1)}} \simeq \frac{U^2 \Pi_0 \xi E}{\mathcal{E}_-(k_\mathrm{F})^2} \simeq \frac{eEa}{U\Pi_0}, \tag{23}$$

with approximating $F_{k\Pi} \simeq \frac{1}{a}$ from Eq. (15). By using the experimental values for the applied electric field $E \simeq 5$ kVm$^{-1}$, $a \simeq 5$ Å and the band gap $U\Pi_0 \simeq 0.1$ eV, the non-reciprocity ratio can be estimated as $\frac{\sigma^{(2)}E}{\sigma^{(1)}} \simeq 2.5 \times 10^{-5}$. This value is consistent with the experimentally observed non-reciprocity ratio $\frac{V_{2\omega}}{V_{1\omega}} \simeq 1 \times 10^{-5}$ with $V_{2\omega} = 5$ μV in Fig. 2**d** and $V_{1\omega} = Ed = 0.5$ V with the distance between the electrodes $d \sim 1$ mm.

**Coupling between polar order and electric field in conventional ferroelectric insulators and narrow-gap polar semiconductors.**

In the case of conventional ferroelectric insulators with wide band gaps, ideally, there are no free electrons to screen the polarization. Consequently, the microscopic polar order *Π* directly corresponds to the macroscopic electric polarization *P*, accompanied by the emergence of surface charges. In such systems, the wide band gap prohibits the inter-band mixing between the valence and conduction bands induced by the external electric field *E*, and the classical $P \cdot E$ coupling dominates. By contrast, in the case of narrow-gap polar semiconductors, even when *Π* appears, the macroscopic electric polarization *P* is screened by mobile electron carriers, leading to drastic reduction of the $P \cdot E$ coupling. Owing to the narrow-gap nature, on the other hand, the external field *E* may cause the inter-band mixing effectively, resulting in the *Π* modulation.

Here, we discuss the difference in the magnitude between the band-mixing-induced coupling $\Pi \cdot E_{\text{eff}}$ and the classical dipolar coupling $P \cdot E$ in narrow-gap polar semiconductors. To address this, we introduce a proportionality parameter $\alpha$ to relate the unscreened and screened electric polarization, $P_{\text{bare}}$ and $P$, respectively: $P = \alpha P_{\text{bare}}$, where $P_{\text{bare}}$ is given by $P_{\text{bare}} = ea\Pi$. For systems with a sufficiently large band gap, $\alpha$ approaches 1, corresponding to a ferroelectric insulator with negligibly small carrier density $n \sim 0$. In the narrow gap case with finite free electrons that screen macroscopic $P$, $\alpha$ can be phenomenologically modelled as $\alpha \simeq \exp(-L/\lambda_{\text{D}})$. Here, $L$ and $\lambda_{\text{D}}$ represent the sample length along the electric field $E$ and the Debye screening length given by $\lambda_{\text{D}} = \sqrt{\varepsilon_r \varepsilon_0 k_B T / n e^2}$, respectively, where $\varepsilon_r$, $\varepsilon_0$, $k_B$, $T$, $n$, and $e$ represent the dielectric constant, the vacuum permittivity, the Boltzmann constant, the temperature, the carrier density, and elementary charge, respectively[50]. For $LuFe_2O_4$, $\varepsilon_r \sim 40$ (refs.[22,51]) and $n \sim 8.9 \times 10^{14}$ $\text{cm}^{-3}$ (ref.[10]), giving the Debye screening length of $\lambda_{\text{D}} = 2.5 \times 10^{-7}$m at 300 K. This yields $\alpha \sim \exp(-8000) \sim 0$, indicating that the macroscopic electric polarization $P$ and the dipolar coupling $P \cdot E$ are vanishingly small for the present sample length. By contrast, the atomic-scale band-mixing-induced coupling $\Pi \cdot E_{\text{eff}}$ remains unscreened. We compare the magnitude of the couplings: $(\Pi E_{\text{eff}})/(PE) = (2V\xi E\Pi)/(\alpha ea\Pi E) \simeq \alpha^{-1}$. $\alpha^{-1} = P_{\text{bare}}/P$ is estimated to be approximately $10^{3475}$ in the present system, showing that the $\Pi \cdot E_{\text{eff}}$ coupling overwhelms the classical dipolar coupling.

**Microscopic perspective on the electronic manipulation of charge ordering.**

We here argue the microscopic perspective on how an applied electric field controls the orientation of the charge-ordered state. In our model, the $Fe^{2+}$-rich and $Fe^{3+}$-rich layers in $LuFe_2O_4$ are described by a simple one-dimensional system with two non-equivalent sites, *A* and *B*, which breaks inversion symmetry (see Extended Data Fig. 2). The system contains localized *d* electrons on each site with different valences that interact with itinerant electrons (referred to as *p* electrons) in the conduction band. We define the primary order parameter, $\Pi$, as the imbalance of the charge occupation $\langle n_{A,B}^{d} \rangle$ of the *d* electrons between the *A* and *B* sites,

$$\Pi = \langle n_A^d \rangle - \langle n_B^d \rangle. \tag{24}$$

This quantity $\Pi$ represents the polar order parameter of the *d*-electron subsystem, and coincides with the charge-order parameter in this model. Note that the $\Pi$ order can be reversed by simply switching the charge occupation $\langle n_{A,B}^{d} \rangle$ between the *A* and *B* sites. This characteristic reflects the situation of $LuFe_2O_4$, where the polar order (charge order) is reversed via electron transfer between the $Fe^{2+}$-rich and $Fe^{3+}$-rich layers without accompanying lattice deformation. In the model, the on-site Coulomb interaction *U* between the *d* and *p* electrons leads to a staggered mean-field potential for *p* electrons at the *A* and *B* sites, which depends on the direction of the *d*-electrons' charge ordering $\Pi$ (see Extended Data Fig. 2). Through this coupling, the charge distribution of the *p* electrons is modified, and the *p* electrons also acquire a microscopic charge polarization, which we parameterize by the variable $\Pi_p$.

A phenomenological free energy describing the *d*-electron order $\Pi$, the *p*-electron polarization $\Pi_p$, and their coupling to an external electric field, is given by

$$F(\Pi, \Pi_p) = F_d(\Pi) + F_p(\Pi_p) + F_{\text{int}}(\Pi, \Pi_p) + F_{\text{field}}(\Pi, \Pi_p). \tag{25}$$

Here, the *d*-electron free energy $F_d(\Pi) = -1/(2\chi_d)\Pi^2 + \beta\Pi^4$ describes the instability leading to the charge order, where the coefficient $1/(2\chi_d)$ changes sign at the charge-ordering temperature $T_{CO}$ to stabilize the spontaneous order $\Pi \neq 0$. The $p$-electron free energy $F_p(\Pi_p) = (1/2\chi_p)\Pi_p^2$ represents the energy cost for the conduction band polarization, where $\chi_p$ is the $p$-band susceptibility. The interaction term $F_{\mathrm{int}}(\Pi, \Pi_p) = -\gamma\Pi\Pi_p$ represents the bilinear coupling between the $d$-electron charge order $\Pi$ and the $p$-electron polarization $\Pi_p$. The coupling $\gamma$ originates from the on-site Coulomb interaction $U$ between the $d$- and $p$-electrons. The field coupling term $F_{\mathrm{field}}(\Pi, \Pi_p)/ea = -\alpha\Pi E - \Pi_p E$ accounts for the interaction with the external electric field $E$, where $e$ and $a$ represent the elementary charge and the distance between the $A$ and $B$ sites, respectively. The term $-\alpha\Pi E$ is the direct dipole coupling allowed by the polar symmetry (the $ABAB$-type charge-ordering pattern in our model), but it is suppressed due to the strong charge screening by the $p$ electrons ($\alpha \ll 1$; see the previous section in Methods). As a result, the external field does not directly control the $d$-electron sector. The dominant interaction pathway is the second term $-\Pi_p E$, which represents the coupling between the $p$-electron polarization $\Pi_p$ and the external electric field $E$ arising from the inter-band hybridization (Fig. 5**a**). We note that time-reversal-odd order parameters (magnetization) are absent in the present nonmagnetic system, excluding any possible magnetoelectric cross terms.

The equilibrium condition $\partial F/\partial \Pi_p = 0$ yields

$$\Pi_p = \chi_p(eaE + \gamma\Pi), \tag{26}$$

showing that $\Pi_p$ depends linearly on both the external field $E$ and the $d$-electron order $\Pi$. This also indicates that modulation of the $p$-electron polarization $\delta\Pi_p$ results in that of the $d$-electron polar order $\delta\Pi = \delta\Pi_p/\chi_p\gamma$. Substituting Eq. (26) back into the total free energy $F(\Pi, \Pi_p)$ [Eq. (25)] allows us to integrate out the $\Pi_p$ degree of freedom and obtain the effective free energy for $\Pi$:

$$F_{\text{eff}}(\Pi) = -V_{\text{eff}}\Pi^2 + \beta\Pi^4 - \tilde{E}_{\text{eff}}\Pi + \text{const}, \tag{27}$$

where $V_{\text{eff}} = (1/2\chi_d) + (\gamma^2\chi_p/2)$ and $\tilde{E}_{\text{eff}} = (\alpha + \gamma\chi_p)eaE$. The above expressions mean that $\Pi$ and $\Pi_p$ are proportional with each other and effectively form a single collective parameter. The influence of the conduction band is fully encoded in the renormalized coefficients $V_{\text{eff}}$ and $\tilde{E}_{\text{eff}}$. This implies that the inter-site Coulomb interaction $V$ in Eqs. (1) and (18) incorporates the contributions from the itinerant electrons; $V = V_{\text{eff}}$. Importantly, although the $d$-electron subsystem is strongly screened from the external field (the net electric polarization $P$ vanishes), the charge order $\Pi$ can still be controlled by the effective field $\tilde{E}_{\text{eff}} \sim \gamma\chi_p eaE$. The effective field $\tilde{E}_{\text{eff}}$ originates from the polarization response of the itinerant electrons and the coupling $\gamma$ induced by the Coulomb interaction $U$ between the $d$- and $p$-electrons.

## Data availability

The data that support the findings of this study are available from the corresponding author upon reasonable request.

## Code availability

The codes used in theoretical simulations and calculations are available from the corresponding authors upon reasonable request.

## Supplementary References

**Acknowledgements**

The authors thank H. Yu, K. Fujiwara, Y. Fukada, N. Inoue, A. Kimura, T. Yamamoto, T. Hioki, and T. Makiuchi for valuable discussions. This work was supported by JST CREST (JPMJCR20C1 and JPMJCR20T2), Grant-in-Aid for Scientific Research (Grants No. JP19H05600, JP23K25816, and JP24K01326), and Grant-in-Aid for Transformative Research Areas (Grants No. JP22H05114 and JP24H02231) from JSPS KAKENHI, MEXT Initiative to Establish Next-generation Novel Integrated Circuits Centers (X-NICS) (Grant No. JPJ011438), Japan, and Sumitomo Chemical.

**Author contribution**

T.K. and Z.C. contributed equally to this work. Z.C. and T.K. constructed the experimental setup, performed the experiments, and collected the data with the help of Y.F., H.A., A.A.K., and Y.H. T.K. and Z.C. analyzed the data with input from Y.F., T.M., Y.O., and E.S. Y.F., T.M., and E.S. developed the theoretical explanation and performed the calculation. N.I. provided the samples. E.S. conceived and supervised the project. T.K. and T.M. wrote the manuscript with review and input from E.S. All the authors discussed the results and commented on the manuscript.

**Competing interests**

The authors declare no competing interests.

**Correspondence** Correspondence and requests for materials should be addressed to T.K. and E.S. (email: kikkawa.takashi@jaea.go.jp, eizi@ap.t.u-tokyo.ac.jp).

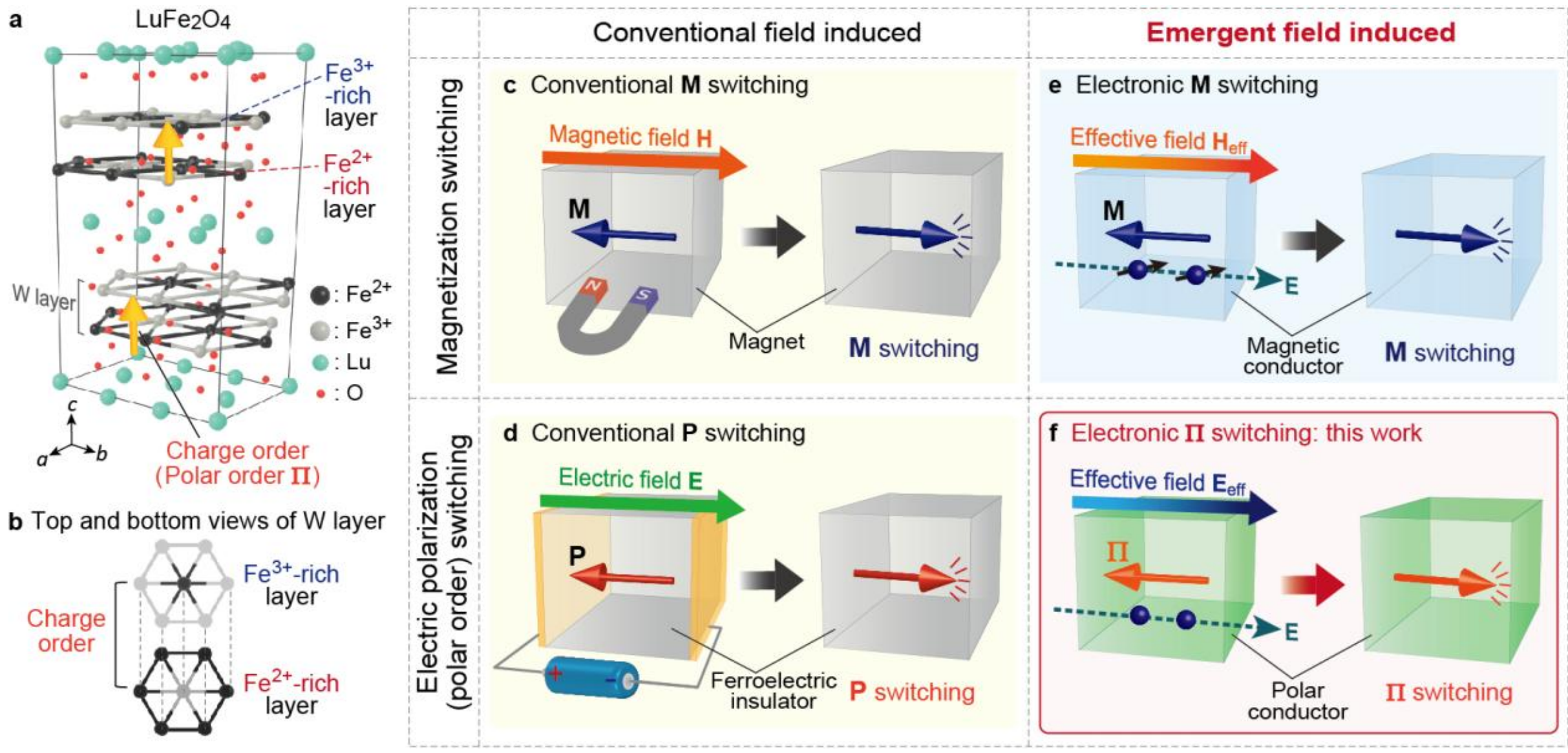

**Fig. 1 | Crystal structure of $LuFe_2O_4$ and concept of magnetic and polar order switching.** **a**, Crystal structure of $LuFe_2O_4$ (ref.[9]), which belongs to the polar monoclinic space group *Cm* at room temperature[12]. The charge-ordering pattern, characterized by the polar order parameter $\mathbf{\Pi}$, appears on the doubly-stacked Fe hexagonal layers, the so-called "W layers", consisting of $Fe^{2+}$(black circles)-rich and $Fe^{3+}$(gray circles)-rich layers along the polar *c*-axis (see Methods and Extended Data Fig. 1 for the X-ray diffraction pattern). **b**, Top and bottom views of the W layer of $LuFe_2O_4$ (each layer is actually stacked, but is depicted with a shift for clarity). **c**, **d**, Conventional field-induced switching of the (**c**) magnetic order (magnetization **M**) in a ferromagnet and (**d**) electric polarization **P** order in a ferroelectric insulator through the external magnetic field **H** and electric field **E**, respectively. Here, the **M** (**P**) switching originates from the classical dipolar coupling $-\mathbf{M}\cdot\mathbf{H}$ ($-\mathbf{P}\cdot\mathbf{E}$). **e**, Electronic-induced **M** switching in a ferromagnetic conductor. Here, an applied field, **E**, generates a non-equilibrium spin polarization, $\delta\mathbf{s}$, which can be regarded as an effective (emergent)

field $\mathbf{H}_{\mathrm{eff}}$ acting on the **M** order, allowing **M** manipulation. **f**, Electronic-induced polar order $\mathbf{\Pi}$ switching in a polar conductor. Here, an applied field, **E**, causes a displacement of the polar order from equilibrium (i.e., non-equilibrium charge polarization), $\delta\mathbf{\Pi}$, which can be regarded as an effective (emergent) field $\mathbf{E}_{\mathrm{eff}}$ acting on the $\mathbf{\Pi}$ order, allowing $\mathbf{\Pi}$ manipulation.

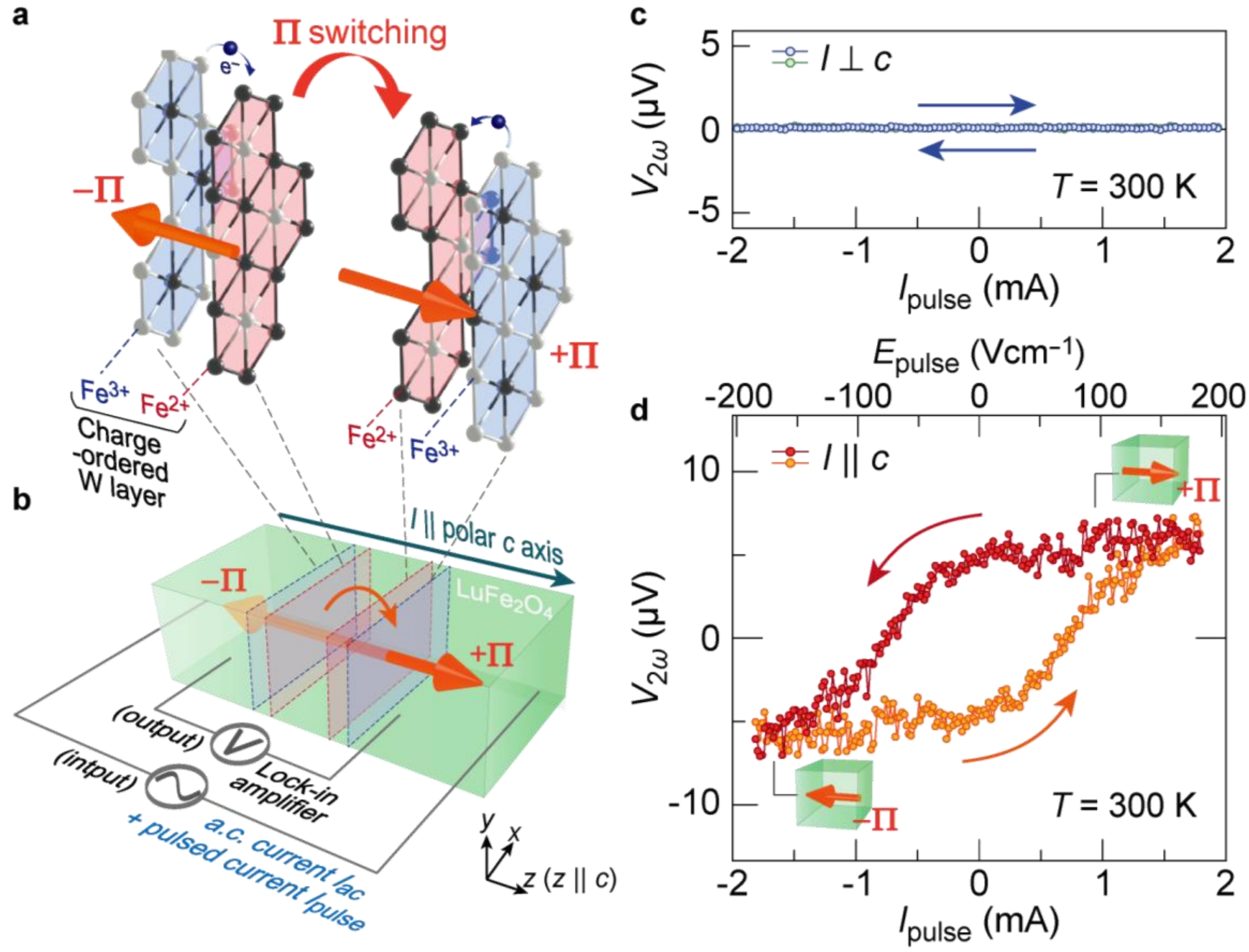

**Fig. 2 | Observation of electronic manipulation of Π order. a**, A schematic illustration of the polar-order (Π) switching in the doubly-stacked Fe hexagonal (W) layers in $LuFe_2O_4$. **b**, A schematic illustration of the experimental setup for the $LuFe_2O_4$ under a four-probe configuration. Square-pulsed and a.c. currents, $I_{pulse}$ and $I_{ac}$ ($= \sqrt{2}I_{\mathrm{rms}}\sin\omega t$), respectively, are applied along the polar *c*-axis (except for the data shown in **c**) and the resultant longitudinal second-harmonic (2*ω*) voltage $V_{2\omega}$ was detected with a lock-in amplifier. **c**, **d**, $I_{pulse}$ dependence of $V_{2\omega}$ in the $LuFe_2O_4$ at $T$ = 300 K under the current $I$ applied perpendicular to the *c* axis ($I \perp c$, blue and green circles shown in **c** for the sample 2) and parallel to the *c* axis ($I \parallel c$, red and orange circles shown in **d** for the sample 1), with the pulse amplitude of $|I_{pulse}| \leq 1.8$ mA, the

pulse width of 80 ms, and the $I_{ac}$ peak-amplitude $\sqrt{2}I_{\mathrm{rms}}$ of 80 μA. $E_{pulse}$ (the upper horizontal axis) in **d** denotes the pulsed electric-field value estimated from $E_{pulse} = (RI_{pulse})/d$, where $R$ and $d$ represent the resistance and the distance between the electrodes, respectively. In **c**, the data plotted as green circles are positioned almost behind the blue circles and therefore not visible in the plot.

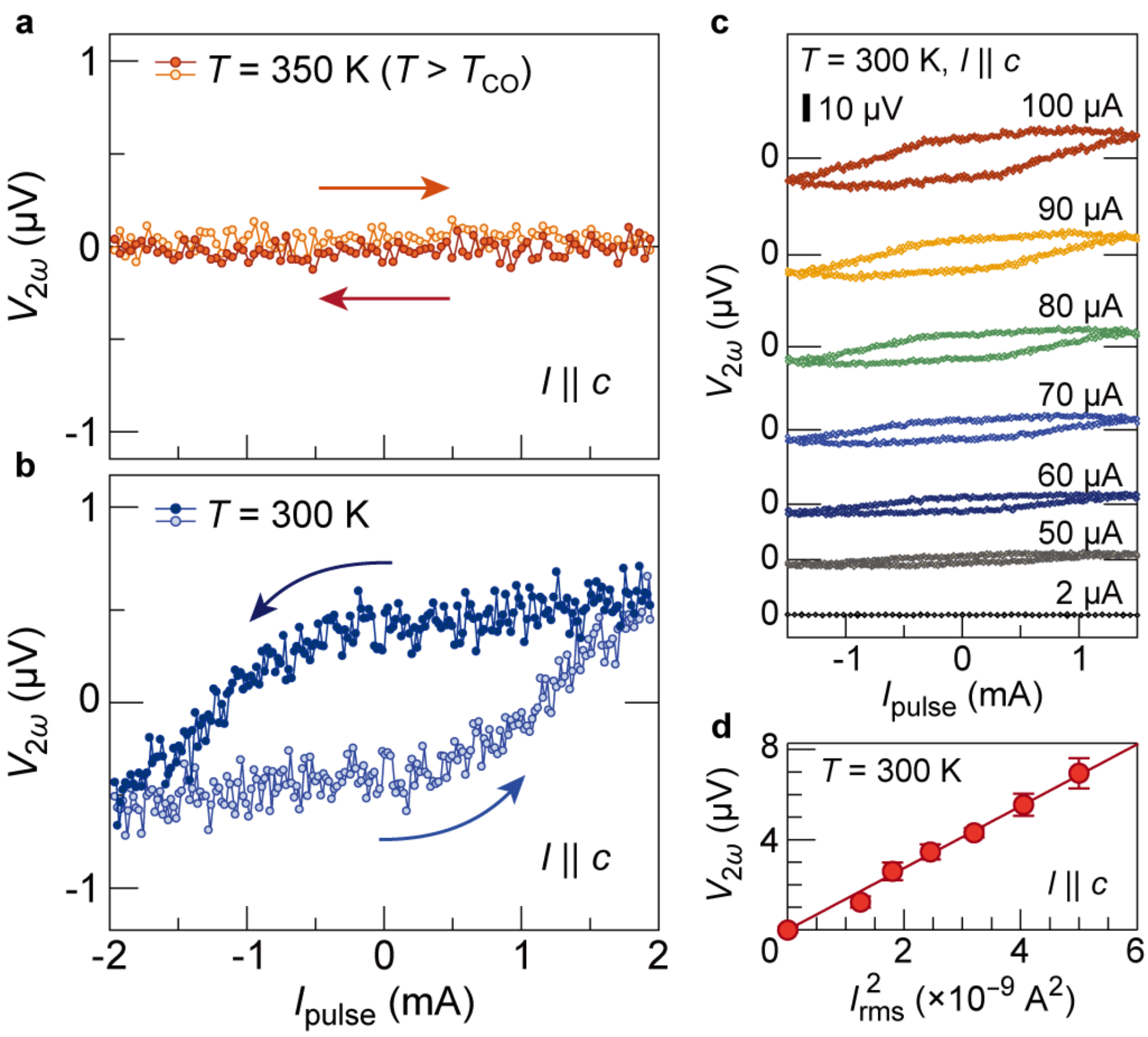

**Fig. 3 | Temperature and a.c. current dependences of nonlinear voltage in $LuFe_2O_4$. a**, **b**, $I_{pulse}$ dependences of $V_{2\omega}$ for the $LuFe_2O_4$ sample 2 at $T$ = 350 K above the three-dimensional charge-ordering temperature $T_{CO}$ = 320 K (red and orange circles shown in **a**) and 300 K (dark blue and light blue circles shown in **b**) under $I \parallel c$. **c**, $I_{pulse}$ dependence of $V_{2\omega}$ for the $LuFe_2O_4$ sample 1 at $T$ = 300 K measured with several $I_{ac}$ peak-amplitudes ($\sqrt{2}I_{rms}$) under $I \parallel c$. The black scale bar represents 10 μV. **d**, $I_{rms}^2$ dependence of the remanent $V_{2\omega}$ at $I_{pulse}$ = 0 mA obtained from the results shown in **c**. The red solid line is a linear fit. The error bar represents the standard deviation (SD).

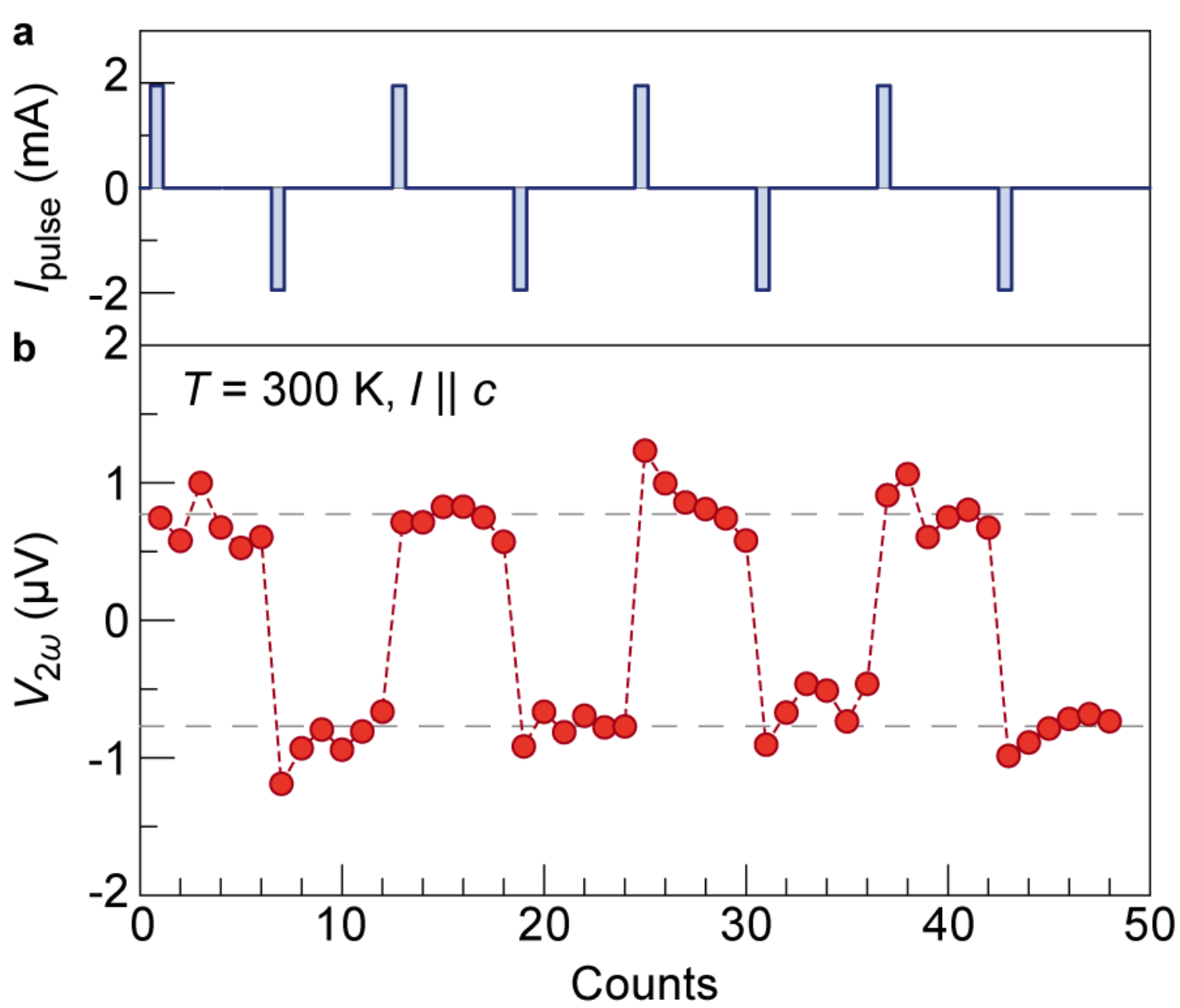

**Fig. 4 | Non-volatile memory operation.** **a**, The input pulsed-current sequence applied along the *c* axis of the $LuFe_2O_4$. Following the initial application of $I_{pulse}$ = + 1.95 mA with the pulse width of 40 ms, the 2*ω* voltage $V_{2\omega}$ is measured after a 45-second interval, with subsequent $V_{2\omega}$ measurements taken five times at 15-second intervals at each point. This process is repeated, while alternating the $I_{pulse}$ direction. **b**, Measured $V_{2\omega}$ for the $LuFe_2O_4$ sample 1 under the a.c. current applied along the *c* axis, with the $I_{ac}$ peak-amplitude $\sqrt{2}I_{\mathrm{rms}}$ of 80 μA, measured at *T* = 300 K.

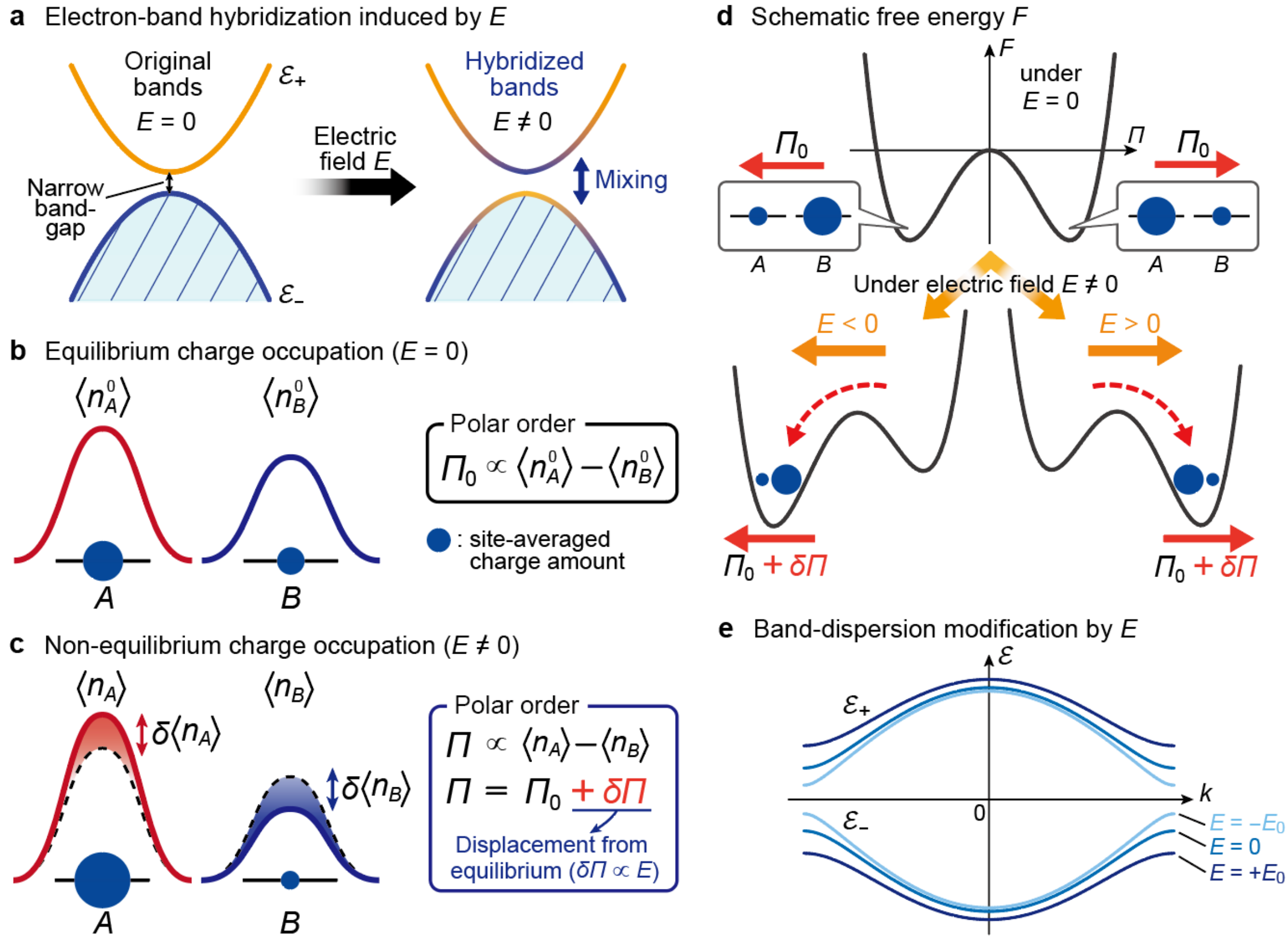

**Fig. 5 | Electronic-induced Π order switching. a**, A schematic illustration of the electric field $E$-induced inter-band hybridization between the two electron bands $\varepsilon_+$ (orange solid curve) and $\varepsilon_-$ (blue solid curve). The hybridization under $E$ is schematically represented by the blending of the band colors. **b**, A schematic illustration of the equilibrium charge occupation in a system comprising two sites (labeled as $A$ and $B$) in a unit cell with the charge ordering characterized by the polar order parameter $\Pi_0$. The red and blue curves depict schematic itinerant-electron distribution at the $A$ and $B$ sites, $\langle n_A^0 \rangle$ and $\langle n_B^0 \rangle$, respectively. Blue circles are depicted at each site for clarity, with their size representing the site-averaged charge amount. **c**, A schematic illustration of the charge-redistribution model via the wave-function mixing

(the first-order Stark effect) upon the application of an electric field *E*, resulting in a modulation of the polar order by $\delta\Pi$, proportional to *E* ($\delta\Pi = \xi E$). **d**, A schematic illustration of the free energy *F* landscape. In the absence of *E*, *F* displays two minima, corresponding to the spontaneous polar order of opposite orientations that are degenerated in energy. The application of *E* leads to a shift of the free energy by $\Delta F_{\mathrm{Coulomb}} = -E_{\mathrm{eff}} \cdot \Pi_0 \propto E$, modulating the two minima of the free energy asymmetrically depending on the *E* direction, enabling the information writing into the $\Pi$ order. **e**, A schematic illustration of the modification of the band dispersions $\mathcal{E}_+$ and $\mathcal{E}_-$ under the electric field $E = \pm E_0$.

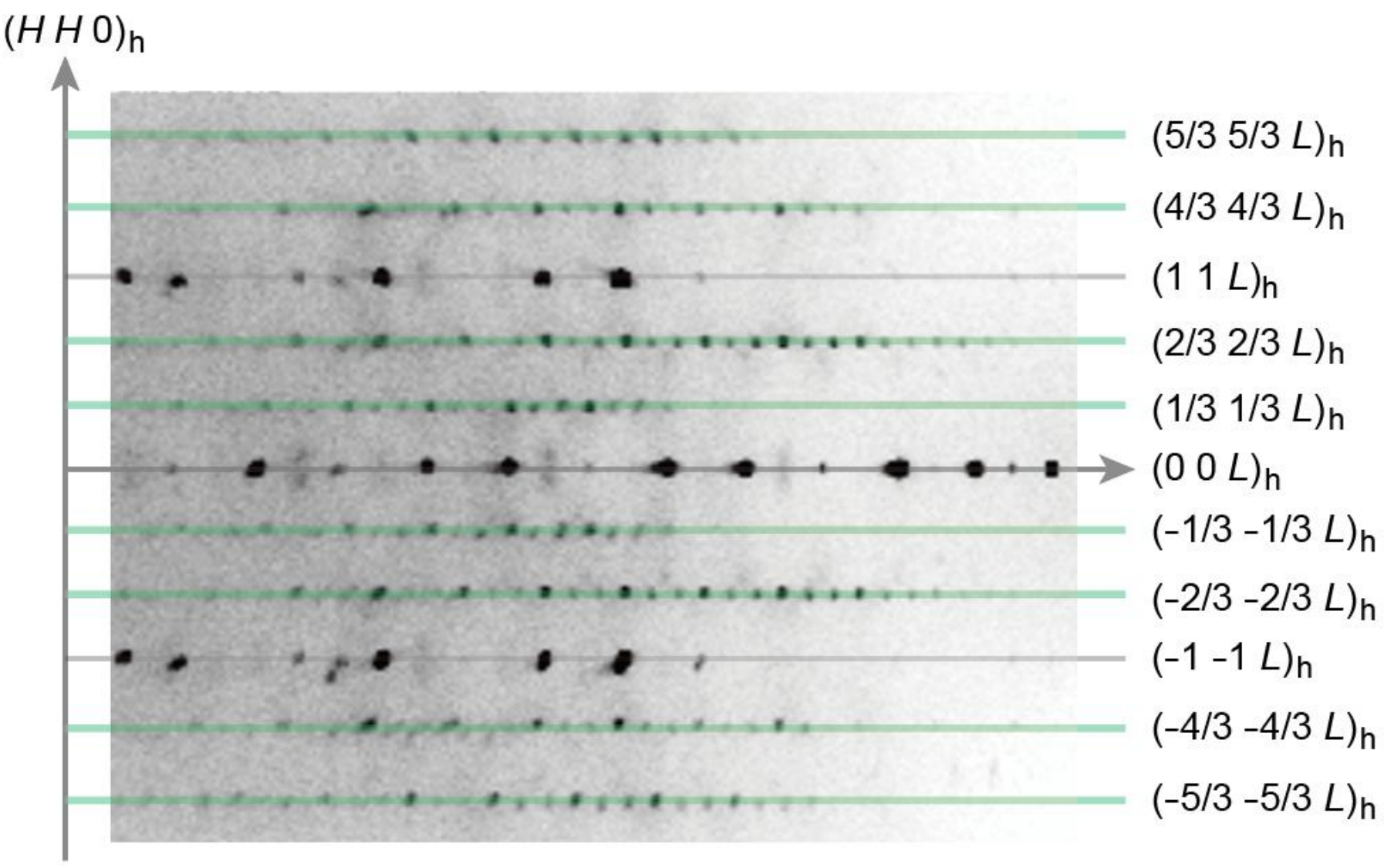

**Extended Data Fig. 1 | X-ray diffraction of $LuFe_2O_4$ at room temperature.** Reciprocal lattice map obtained by X-ray oscillation photography using a Cu *Kα* radiation and an imaging-plate detector. Diffraction spots on the green markers, such as $(1/3\ 1/3\ L)_h$ and $(2/3\ 2/3\ L)_h$, where $L = 0.5n$ ($n$: integer), correspond to superlattice reflections arising from three-dimensional charge ordering.

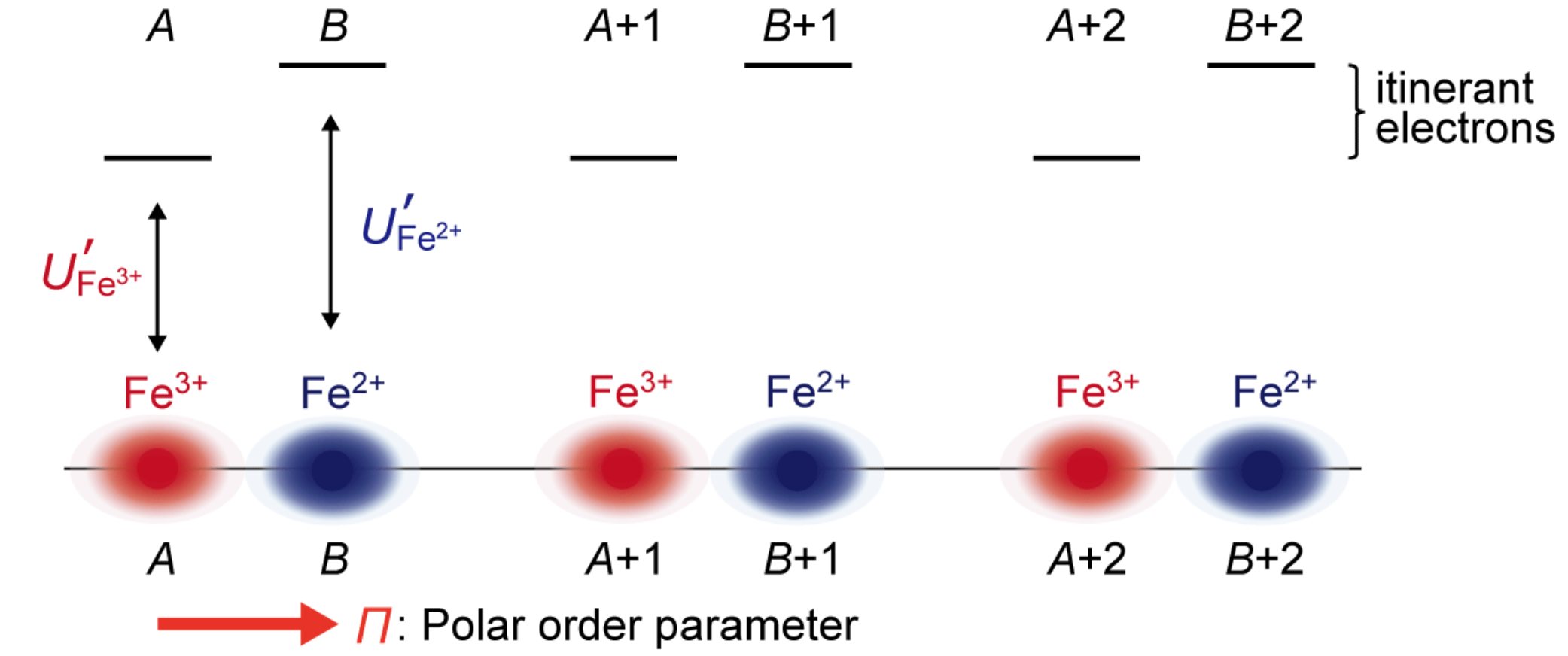

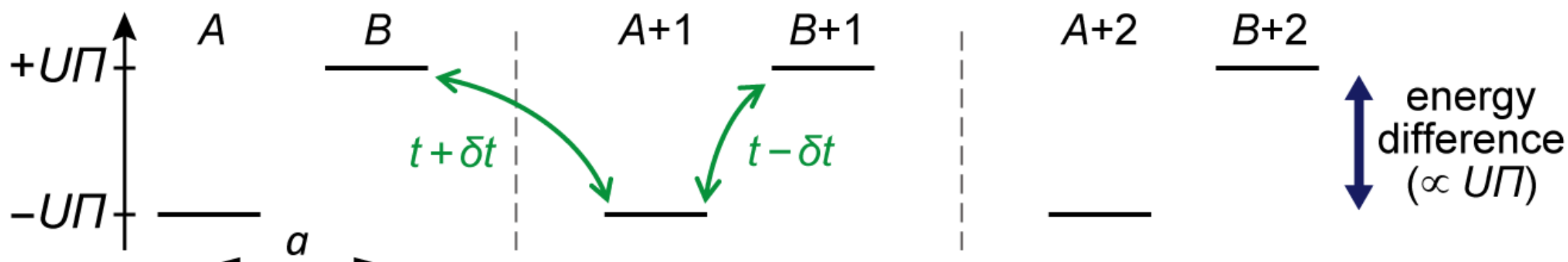

**Extended Data Fig. 2 | Charge-ordering model and one-electron energy diagram for a two-site system with broken inversion symmetry. a**, A schematic illustration of an electron system comprising two sites in a unit cell, labeled as *A* and *B*, separated by the distance $a$. The red and blue clouds depict the schematic *d* electrons of $Fe^{3+}$ and $Fe^{2+}$ ions at the *A* and *B* sites, respectively. $U'_{\mathrm{Fe}^{3+}}$ ($U'_{\mathrm{Fe}^{2+}}$) represents the on-site Coulomb energy between an itinerant electron and the *d* electrons of the $Fe^{3+}$ ($Fe^{2+}$) ion. Because $U'_{\mathrm{Fe}^{3+}}$ and $U'_{\mathrm{Fe}^{2+}}$ are different, the mean-field energy level of the itinerant-electron orbitals varies depending on whether the $Fe^{3+}$ or $Fe^{2+}$ ion is located at the *A* or *B* sites. **b**, A modeled energy diagram for the itinerant-electron orbital, wherein the energy difference is given by the product of $\Pi$ $(-1 \leq \Pi \leq -1)$ and the on-site Coulomb energy *U*. $t \pm \delta t$ represents the alternating hopping parameters that differ from each other to account for the broken inversion symmetry. Upon the

application of *E*, $U\Pi$ may change to $U(\Pi + \delta\Pi)$ (Hartree correction, with $\delta\Pi \propto E$). This mechanism alters the curvature of the electron band $\varepsilon_{\pm}(k)$ (hence, group velocity) depending on the *E* direction (Fig. 5**e**).